\documentstyle[multicol,aps,epsf]{revtex}
\begin{document}
\draft
\title{Gamow-Teller Strengths of  the Inverse-Beta Transition  
$^{176}$Yb${\rightarrow}^{176}$Lu for Spectroscopy of 
Proton-Proton and other sub-MeV Solar Neutrinos }

\author{  M.~Fujiwara,$^{1,2}$ H.~Akimune,$^{3}$ A.M.~van~den~Berg,$^{4}$
M.~Cribier,$^{5}$ I.~Daito,$^{6}$
H.~Ejiri,$^{1}$ H.~Fujimura,$^{1}$ Y.~Fujita,$^{6}$  C.D.~Goodman,$^{8}$
K.~Hara,$^{1}$ M.N.~Harakeh,$^{4}$ F.~Ihara,$^{1}$ T. Ishikawa,$^{9}$
J.~J\"anecke,$^{10}$ T.~Kawabata,$^{9}$ R.S.~Raghavan,$^{11}$
K.~Schwarz,$^{1}$ M.~Tanaka,$^{12}$ 
T.~Yamanaka,$^{1}$ M.~Yosoi,$^{1}$ and
R.G.T.~Zegers$^{2,4}$ }

\address{
$^{1}$Research Center for Nuclear Physics, Osaka University,
              Mihogaoka 10-1, Ibaraki, Osaka 657-0047, Japan  \\
$^{2}$Advanced Science Research Center, JAERI, Tokai, Ibaraki, 319-1195
Japan \\
$^{3}$Konan University, Higashinada, Kobe 658, Japan \\
$^{4}$Kernfysisch Versneller Instituut, 9747 AA Groningen,
                The Netherlands \\
$^{5}$CEA/Saclay, DAPNIA/SPP 91191 Gif-sur-Yvette Cedex, France  \\
$^{6}$Nagoya University, Department of Physics, 464 Nagoya, Japan \\
$^{7}$Department of Physics, Osaka University, 560-0043, Japan \\
$^{8}$Indiana University Cyclotron Facility, Bloomington, Indiana, USA \\
$^{9}$Department of Physics, Kyoto University, Kyoto 606-8502, Japan\\
$^{10}$Department of Physics, University of Michigan, Ann Arbor, MI
                48109-1120, USA \\
$^{11}$Bell Laboratories, Lucent Technologies, Murray Hill, New Jersey
07974, U.S.A. \\
$^{12}$Kobe Tokiwa Jr. College, Nagata 653, Japan  \\ }

\date{\today}
\maketitle
\begin{abstract}
Discrete Gamow-Teller (GT) transitions, $^{176}$Yb${\rightarrow}^{176}$Lu 
at low excitation energies have been measured via 
the ($^3$He,$t$)  reaction at 450 MeV and at 0$^{\circ}$. 
For $^{176}$Yb, 
two low-lying states are observed, setting low thresholds 
Q(${\nu}$)=301 and 445 keV for neutrino (${\nu}$) capture. 
Capture rates estimated from the measured GT strengths, 
the simple two-state excitation structure, and the low Q(${\nu}$) in 
Yb--Lu indicate that Yb-based ${\nu}$-detectors are well suited for 
a direct measurement of the complete 
sub-MeV solar electron-neutrino (${\nu}_e$) spectrum 
(including $pp$ neutrinos) where definitive effects of 
flavor conversion are expected.
\end{abstract}

\begin{multicols}{2} 
The major result of experimental solar-neutrino research to date shows  
that  the observed fluxes of solar neutrinos are much reduced 
compared with  theoretical predictions based on the standard solar model (SSM) 
\cite{Bahcall98}.  
Particularly sharp questions are posed by the results 
from the Gallex \cite{Hampel96} and Sage \cite{Gavrin99} experiments at 
the low-energy end of the spectrum (from the $p+p$, $^7$Be and 
CNO reactions in the sun) that are in contrast with those 
from Super-Kamiokande at high energies 
(from the decay of $^8$B)  \cite{suzuki98}. 
The low-energy results  imply a  negligible flux of $^7$Be neutrinos which 
is inconsistent with the observed sizeable flux of high-energy $^8$B 
neutrinos  because the $^8$B activity in the sun cannot arise without the
precursor $^7$Be in the reaction $^7$Be+$p{\rightarrow}^8$B.  
These  results suggest mechanisms beyond 
possible astrophysical shortcomings of the SSM. 
The only other possibility  is non-standard neutrino physics, viz. 
the conversion of the original electron-flavor solar neutrinos (${\nu}_e$) to 
undetected ${\mu}$ and  ${\tau}$ flavors. 
Evidence for ${\nu}_{\mu}$--${\nu}_{\tau}$ flavor conversion has recently been 
observed by Super-Kamiokande \cite{kajita98}. 

The conceivable conversion mechanisms produce ${\nu}_e$ flux deficits 
that are energy dependent (typically strongest at low energies) and 
result in large  deficits of either the $pp$ and/or 
the $^7$Be ${\nu}_e$   fluxes, 
the two dominant features of the sub-MeV spectrum. 
A deficit in the model-independent $pp$-neutrino flux and/or 
a $^7$Be   ${\nu}_e$ flux measurably smaller than the total
all-flavor flux (expected from the Borexino experiment \cite{Bx} 
in particular),  would provide the most direct proof of flavor conversion. 
A real-time measurement of the complete ${\nu}_e$  spectrum from the sun and  
source-specific fluxes is, thus, of central interest for solving the 
solar-${\nu}$ problem. 
Such data are not available yet since the  
operating low-energy Ga  detectors yield only the integral 
signal rate above a threshold, not the fluxes
from specific solar  ${\nu}$ sources. 

Recently, Raghavan \cite{Raghavan97} suggested a possible way to 
construct a low-threshold, flavor-specific scheme for real-time detection 
of solar neutrinos via neutrino captures based on the  charged 
current mediated Gamow-Teller (GT) transitions 
${\nu}_e + ^{176}$Yb${\rightarrow} e^-$ + $^{176}$Lu$^*$  and 
${\nu}_e + ^{160}$Gd${\rightarrow} e^-$ + $^{160}$Tb$^*$. 
The basic idea is to identify absorption events of  $pp$ and  
$^7$Be neutrinos by a delayed coincidence between a
prompt $e^-$ event and a cascade ${\gamma}$-ray via  isomeric 
states in $^{176}$Lu or $^{160}$Tb. 
The coincidence tag (with time gates ${\leq}$10$^{-7}$ s) allows 
suppression of background events  by a factor of 10$^{7}$.  
For the first time, 
this scheme offers the tools for practical real-time spectroscopy 
even of  ${pp}$--neutrinos despite
the formidable backgrounds that have precluded low-energy   ${\nu}$
spectroscopy so far. 
Feasibility studies are in
progress for constructing such a solar ${\nu}$ detector 
LENS (Low-Energy Neutrino Spectroscopy) \cite{webpage}. 

The basic input data for designing LENS  are the cross sections of 
the GT transitions in Yb and Gd  for which the weak matrix elements 
B(GT) must be determined for each of  the low-lying 1$^+$ states excited 
by neutrino capture. 
The low energy-excitations in Yb-Lu and Gd-Tb consist of several 
close-lying known 1$^+$ states. 
Clarification of the 
states relevant to LENS thus requires high-resolution spectra. 
We have therefore measured  high resolution ($^3$He,$t$) spectra at 0$^{\circ}$
using a $^3$He beam at 450 MeV. 
This work complements a different approach to find B(GT) via ($p,n$) 
measurements \cite{Bhattacharya} on the same targets. 

\begin{figure}
\begin{center}
\begin{minipage}[t]{6.5cm}
\epsfysize=6.5cm
  \epsffile{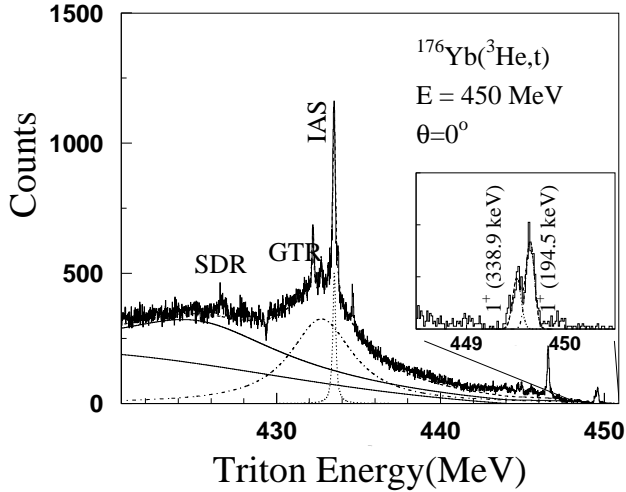}
\end{minipage}
\begin{minipage}[t]{8.0cm}
\caption{
 Triton energy spectrum from the
$^{176}$Yb($^3$He,$t$) reaction at 0$^{\circ}$ taken with a $^3$He beam
of 150 A${\cdot}$MeV. A sharp IAS peak and  two broad bumps corresponding to
the Gamow-Teller resonance (GTR), and the spin dipole resonance (SDR) 
are observed. The global structure of the spectrum is fitted by 
assuming an underlying continuum due to quasi-free scattering. 
Discrete states at the low excitation energy are displayed in the 
expanded inset. Sharp peaks near to IAS are due to C and O contaminations }
\label{fig-yb}
\end{minipage}
\end{center}
\end{figure}

The ($^3$He,$t$) reaction at 450 MeV 
has already been  demonstrated to be useful in getting the 
B(GT) values. Since the full Fermi transition strength B(F)=N--Z is exhausted
by the isobaric analog state (IAS), the B(GT) values can be
obtained from the  singles ($^3$He,$t$) spectra  by using 
the ratio of the cross sections to the 1$^+$ states and the 
IAS \cite{fujiwara96}.  The measurements described below show that 
the  B(GT) values obtained here can be used reliably to estimate solar
neutrino rates in LENS.

The ($^3$He,$t$) experiment was performed at E($^3$He)=450 MeV
at the RCNP cyclotron facility of Osaka University using the 
spectrometer Grand Raiden \cite{fujiwara99} which was set at 
$\theta$ = 0$^{\circ}$ with a solid angle of ${\Delta}{\Omega}$ = 1.6  msr. 
Both the incident $^{3}$He$^{++}$ beam and the reaction products entered into
the spectrometer set at 0$^{\circ}$. 
A metallic target of $^{176}$Yb (96.7 \% enrichment, 3.3 mg/cm$^2$) was used.
The $^{3}$He$^{++}$ beam, which has a much small magnetic rigidity than 
the tritons,was stopped at the internal Faraday cup in the first
dipole (D1) magnet. 
Details of the experimental setup are described in
Ref. \cite{fujiwara99}. 
Particles analyzed by the spectrometer  were detected
in  the focal-plane detection system consisting of two
multi--wire drift chambers (MWDC's) and two ${\Delta}$E plastic
scintillation counters, whose signals were used for particle
identification as well as for triggering the events.
The MWDC's gave information
on the arrival positions at the focal plane and the scattering
angles of the tritons.

Very good identification
with the focal-plane detector of the spectrograph was required to overcome the
huge background due to the  $^3$He$^{++}$ ${\rightarrow}$  $^3$He$^{+}$
atomic charge-exchange process \cite{Dennis}, since the Q-values for the
($^3$He,$t$) reactions on $^{176}$Yb leading to
the lowest 1$^+$ states in the residual nuclei is ${\leq}$ 0.3 MeV.
For this purpose, we added a plastic
scintillation counter with a thickness of 1 mm in front of the
two ${\Delta}E$ counters of 10 mm thickness for improved
particle identification.  
This front counter should be very thin
to avoid creating tritons induced  by ($^3$He,$t$) reactions in
the scintillator itself. With this technique, we could obtain
 a spectrum  free from the
$^3$He$^{++}$ ${\rightarrow}$  $^3$He$^{+}$ atomic charge-exchange
process.

To achieve a high energy resolution of around 100  keV,  we installed
an energy defining slit in the beam-transport system from the K=400 MeV 
ring cyclotron to the target position.  
The image-defining slit was set at the first beam-focusing position 
of the beam-line.
The technique of dispersion matching between the beam-line and
the spectrometer was employed to obtain the required high resolution.
An overall energy resolution of ${\Delta}E$ = 100${\sim}$140 keV was 
obtained, which was sufficient to resolve the low-lying 1$^+$ states  
in $^{176}$Lu.
\begin{table}
\caption{Gamow-Teller and Fermi strengths B(GT) and B(F) of the levels 
observed in the  
$^{176}$Yb($^3$He,$t$)$^{176}$Lu and  $^{160}$Gd($^3$He,$t$)$^{160}$Tb 
reactions.}
\begin{center}
\begin{tabular}{ccc}
\multicolumn{3}{c}{
$^{176}_{~70}$Yb$_{106}$($^3$He,$t$)$^{176}_{~71}$Lu$_{105}$  }            \\
\hline
J$^{\pi}$ & $E_x$(keV)  & B(GT) or B(F) \\
\hline
1$^+_1$     &  194.5     & 0.20${\pm}$0.04           \\
1$^+_2$     &  338.9    & 0.11${\pm}$0.02                \\
1$^+_3$     &  3070       & 0.62${\pm}$0.08           \\
0$^+_{IAS}$     &  16026 ${\pm}$ 6  &  36 \\
\hline
\multicolumn{3}{c}{
 $^{160}_{~64}$Gd$_{96}(^3$He,$t$)$^{160}_{~65}$Tb$_{95}$ }    \\
\hline
J$^{\pi}$ & $E_x$(keV)   & B(GT) or B(F) \\
\hline
1$^+_1$     &  138.7              &  0.054${\pm}$0.009     \\
1$^+_2$     &  232.8         &    0.014${\pm}$0.002        \\
1$^+_3$     &  478.2             &   0.160${\pm}$0.03      \\
1$^+_4$     &  573.0              & 0.021${\pm}$0.004  \\
1$^+_5$     &  664.7            & 0.031${\pm}$0.005  \\
1$^+_6$     &  1120              & ${\sim}$0.034   \\
1$^+_7$     &  1310               & ${\sim}$0.025     \\
1$^+_8$     &  1500                  & ${\sim}$0.043       \\
1$^+_9$     &  1610                 & ${\sim}$0.034     \\
1$^+_{10}$     &  1670                    &  ${\sim}$0.051        \\
1$^+_{11}$     &  1750                    & ${\sim}$0.051       \\
1$^+_{12}$     &  1900                  & ${\sim}$0.056      \\
1$^+_{13}$     &  2040                   & ${\sim}$0.259  \\
0$^+_{IAS}$     &  15019 ${\pm}$ 6      &  32 \\
\end{tabular}
\end{center}
\label{Yb-Gd}
\end{table}

To confirm the validity of the particle identification, 
we performed an additional experiment where the  $^{176}$Yb target
was backed with a thin mylar foil (0.086 mg/cm$^2$) to reduce the huge
$^3$He$^{++}$ ${\rightarrow}$  $^3$He$^{+}$  yields.
In this run, a high-resolution measurement was not possible 
because of the achromatic beam-transport mode. 
However, it was confirmed that the unresolved yield  for
the 1$^+$ states at 195 keV and 339 keV has the same relative intensity as
the summed yield for these two states in the 
0$^{\circ}$ $^{176}$Yb($^3$He,$t$) spectrum taken in the 
dispersion-matching high-resolution mode.

To gauge  the B(GT) values obtained in the ($^{3}$He,$t$) analysis, a 
calibration measurement was performed using the neighboring  
$^{164}$Dy($^{3}$He,$t$) reaction, leading to the 1$^+$ ground state 
in $^{164}$Ho. The $^{164}$Ho ground state is known to 
${\beta}$-decay to the $^{164}$Dy ground  state with a log {\it ft} value 
of 4.6 (Ref.~\cite{NDS}). This yields a B(GT) value of 
0.293${\pm}$0.006 for the Dy-Ho transition. The $^{164}$Dy result  
can thus be used to independently calibrate the strong-interaction part of 
the ($^3$He,$t$) reaction cross-sections for measurements on the 1$^+$ states.

Fig.~\ref{fig-yb} shows the ($^3$He,$t$) spectrum on
$^{176}$Yb  at ${\theta}$=0$^{\circ}$.
We deduced the relative transition strengths of  the 194.5 keV to the 
338.9 keV states, by fitting the peaks with a Gaussian shape 
(see Fig.~\ref{fig-yb}), to be 1.0:0.55.
The transition strengths to the excited states were compared
 to that of the IAS. The excitation strengths to GT states and
the IAS depend, however, on  the volume
integrals of the central parts of the effective interaction, $J_{\tau}$
and $J_{{\sigma}{\tau}}$, and distortion effects parameterized
by  $N^D_{\tau}$ and $N^D_{{\sigma}{\tau}}$ \cite{goodman80}.
Using the observed strengths to the ground 1$^+$ state and the IAS in the
$^{164}$Dy($^3$He,$t$)$^{164}$Ho reaction at the same bombarding energy of
450 MeV, we calibrated the ratio of the interaction strengths including
the distortion effects as
\begin{equation}
\frac{N^D_{\sigma\tau}{\mid}{J_{\sigma\tau}}{\mid}^2}{N^D_{\tau}{\mid}{J_{\tau}}{\mid}^2} =\frac{(\frac{d{\sigma}}{d{\Omega}})_{GT}B(F)}{(\frac{d{\sigma}}{d{\Omega}})_{IAS}B(GT)}= 7.44{\pm}0.79.
\end{equation}
Here, we used B(GT)=0.293${\pm}$0.006, B(F)=N--Z=32, and
$(\frac{d{\sigma}}{d{\Omega}})_{GT}/(\frac{d{\sigma}}{d{\Omega}})_{IAS}$=0.068${\pm}$0.007 for the transition to  the ground state of $^{164}$Ho. The value of
$7.44{\pm}0.79$ for the ($^3$He,$t$) reaction at 150 A${\cdot}$MeV
agrees nicely with the values obtained by the
scaling relation $[(E/A)/55 MeV]^2$ which fits the systematics of  
(p,n) reaction data \cite{goodman94}, and has been verified 
specifically for the $^{176}$Yb(p,n) reactions with values 
of 4.76 obtained for E$_p$=120 MeV
and 8.46 for 160 MeV \cite{Bhattacharya}. 

The present calibration method is applicable since the distortion effect
and the ratio of the effective interactions
${\mid}J_{\sigma\tau}/J_{\tau}{\mid}$ are
expected to be the same as those for the $^{176}$Yb($^3$He,$t$)
reaction and the kinematic factors are cancelled out. Empirically,
the volume integrals of the effective
interaction and the distortion effect are not significantly 
different among target nuclei in the same mass region.

The B(GT) values for the 1$^+$ states in $^{176}$Lu  were
obtained using the equation
\begin{equation}
B(GT) = B(F) \frac{1}{7.44}
\frac{(\frac{d{\sigma}}{d{\Omega}})_{GT}}{(\frac{d{\sigma}}{d{\Omega}})_{IAS}},
\end{equation}
where B(F) is (N--Z), and the ratio of
$(\frac{d{\sigma}}{d{\Omega}})_{GT}/(\frac{d{\sigma}}{d{\Omega}})_{IAS}$ is
obtained from the ($^3$He,$t$) spectra at 0$^{\circ}$.
The  B(GT) values deduced from the 0$^\circ$ cross sections measured in
 the ($^3$He,$t$) experiment at 450 MeV are listed  in
Table~\ref{Yb-Gd}. The B(GT) values obtained for the 
possible low-lying 1$^+$ levels in $^{160}$Tb from 
$^{160}$Gd($^3$He,$t$)$^{160}$Tb  are also listed 
for reference.  The B(GT) values for $^{176}$Yb obtained here are in 
agreement with  the (p,n) results \cite{Bhattacharya}.

\begin{figure}
\begin{center}
\begin{minipage}[t]{5.5cm}
\epsfysize=5.5cm
  \epsffile{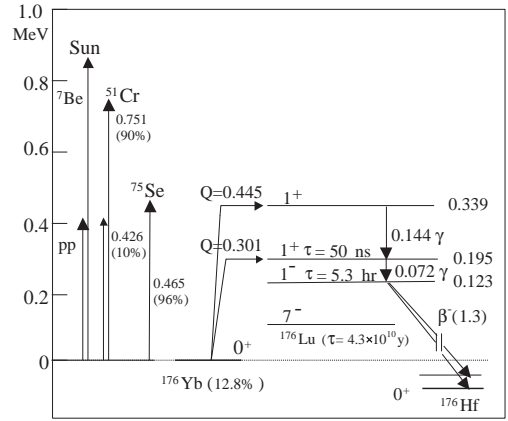}
\end{minipage}
\begin{minipage}[t]{8.0cm}
\caption{Level scheme and  ${\gamma}$-ray tags for solar neutrino detection 
by $^{176}$Yb. All energies are in MeV.
}
\label{level}
\end{minipage}
\end{center}
\end{figure}

The reliability of the B(GT) values obtained here is supported 
by the following considerations:
1) scaling of reaction cross sections for GT resonances relative to the
Fermi strength observed in the IAS, 2) direct use of the known $^{164}$Dy weak 
matrix element to calibrate the strong-interaction factor, 
3) agreement of the deduced strong interaction factor
with the energy scaling systematics of (p,n) reactions in general 
and those for $^{176}$Yb in particular, and 4) agreement of 
the B(GT) values obtained in the ($^3$He,$t$) and (p,n) works for $^{176}$Yb 
and $^{160}$Gd.

The impact of the B(GT) values for $^{176}$Yb  may now 
be examined. The complete level scheme
involved in an Yb-based detector is shown in Fig.~\ref{level}. 
These data and those of Table~\ref{Yb-Gd} show that  only the two
1$^+$ levels at 194.5 keV and 338.9 keV in the final nucleus $^{176}$Lu 
are populated by ${\nu}$ capture below 3 MeV. 
The thresholds for ${\nu}$ capture are determined by the 
level energies and the $^{176}$Yb-$^{176}$Lu mass difference 
as Q(${\nu}$) = 301 keV and 445 keV
for the above two states. 
The Q value of 301 keV lies below 426 keV, 
the endpoint of the $pp$ ${\nu}$ continuum.
The B(GT) value for this transition is the larger of the two. 

\begin{figure}
\begin{center}
\begin{minipage}[t]{7.0cm}
\epsfysize=7.0cm
  \epsffile{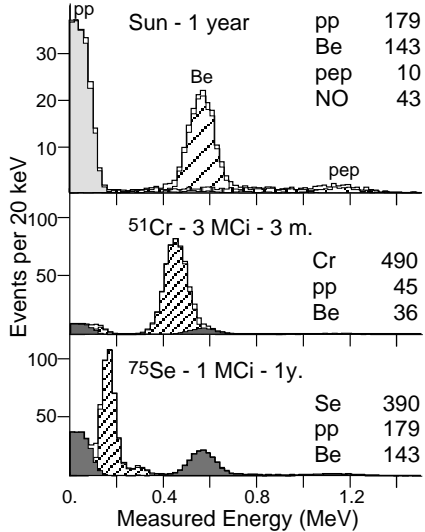}
\end{minipage}
\begin{minipage}[t]{8.0cm}
\caption{Calculated  ${\nu}_e$ spectra in a Yb-doped scintillator with 
a mass of 20 tons of natural Yb exposed by ${\nu}$'s from a) the sun, 
  b) a radioactive source of  $^{51}$Cr, and c) a source of $^{75}$Se.
}
\label{Mont}
\end{minipage}
\end{center}
\end{figure}

The above data on the ${\nu}$-capture level structure and the capture 
strengths establish the fundamental suitability of Yb-LENS for the 
long-standing problem of not only the detection but also  the spectroscopy 
of  solar neutrinos. The response of Yb-LENS to solar neutrinos, 
calculated for a 20 ton natural Yb target  with fluxes given by 
the SSM \cite{Bahcall98} is shown in Fig.~\ref{Mont}a. 
The observable spectrum 
is that of the prompt electron $e^-$  with the 
energy $E_{e^-} = E_{\nu}$ - Q$_{\nu}$ that follows ${\nu}$-capture. 
Consequently, the incident ${\nu}_e$ spectrum is  recorded directly. 
The principal features of 
the $pp$, $^7$Be, $pep$ and the underlying continuum from 
CNO  reactions in the sun are clearly resolved assuming a conservative 
1 ${\sigma}$  energy resolution of ${\Delta}E/E{\sim}7\%/\sqrt{E(MeV)}$.

Because of the importance of absolute fluxes for the three major 
${\nu}$ sources [$pp$(E$_{max}$=420 keV),
$^7$Be (862 keV)  and $pep$ (1442 keV)] B(GT) values must be known 
individually for the two states in $^{176}$Lu. 
The simplicity of the level structure of  $^{176}$Lu (only 2 states 
below 3 MeV) is valuable in contrast to the case of $^{160}$Gd 
(see Table~\ref{Yb-Gd}) in which  5 levels below 1 MeV participate 
and more than 13 levels below ${\sim}$2 MeV. 
Therefore, the use of Gd is less favorable than Yb since interpretation 
of the experimental data  is more complicated besides 
presenting severe problems in the practical operation of the detector.  

The merits of $^{176}$Yb are reinforced by plans to measure the 
${\nu}$-absorption cross sections directly using mono-energetic ${\nu}$'s
from megaCurie (MCi) radioactive sources. 
Such a calibration requires a series of sources, each with a single 
neutrino line, selected to discriminate level thresholds so that 
cross sections to individual levels can be deduced at least from the 
combined data. In principle,  as many sources (lines) are needed 
as target levels revealed in this work.
Clearly, such measurements are  practically ruled out for Gd 
with so many levels to characterize. For the two levels in $^{176}$Yb, 
two sources are proposed. The first, $^{51}$Cr \cite{Raghavan78,Cribier96} 
produces a 752 keV line which populates both the Yb transitions 
(the weaker line at  426 keV produces a much weaker signal). 
The second source, $^{75}$Se \cite{Cavaignac,Kornukhov} produces 
a line at 465 keV which is sensitive only to the lower level. 
Figs.~\ref{Mont}b and c depict  the LENS response for these two calibration 
lines.
They show that the neutrino response of an Yb-LENS can be calibrated 
experimentally without interference from a ``background'' of solar neutrinos.

In summary, we report experimental results on the Gamow-Teller
transitions from $^{176}$Yb to $^{176}$Lu using the ($^{3}$He,$t$) reaction at
$E$($^{3}$He)=450 MeV. A high-resolution measurement enabled us to resolve
the low-lying $1^+$ states in the residual nucleus.
By taking the ratios of the excitation strengths for the IAS and the
Gamow-Teller states, we deduce the B(GT) values for low-lying $1^+$ states
after calibrating the reaction mechanism. The data establish the foundations 
for a practical real-time $^{176}$Yb detector for sub-MeV solar ${\nu}$'s 
including $pp$ ${\nu}$'s. Solar ${\nu}$ absorption 
rates for a 20-ton Yb detector are given together with those for
the $^{51}$Cr and $^{75}$Se ${\nu}$ calibrations.

The authors acknowledge the RCNP cyclotron staff for their support during
the experiment. The present work has been  supported by the Ministry of
Education, Science, Sports and Culture (Monbusho) with Grant No. 09041108
and by the Japan Society for the
Promotion of Science (JSPS).


\end{multicols}
\end{document}